# Direct-Current Generator Based on Dynamic Water-Semiconductor Junction with Polarized Water as Moving Dielectric Medium

*Yanghua Lu[1#], Yanfei Yan[1#], Xutao Yu[1#], Xu Zhou[2#], Sirui Feng[1], Chi Xu[1], Haonan Zheng[1], Zunshan Yang[1], Linjun Li[3], Kaihui Liu[2] and Shisheng Lin[1,3*]*

[1]College of microelectronics, College of Information Science and Electronic Engineering, Zhejiang University, Hangzhou, 310027, P. R. China

[2]State Key Lab for Mesoscopic Physics and Frontiers Science Center for Nano-optoelectronics, Collaborative Innovation Center of Quantum Matter, School of Physics, Peking University, Beijing, 100871, P. R. China

[3]State Key Laboratory of Modern Optical Instrumentation, College of Optical Science and Engineering, Zhejiang University, Hangzhou, 310027, P. R. China

\# These authors contributed equally to this work

[*]Correspondence: shishenglin@zju.edu.cn.

## Abstract

There is a rising prospective in harvesting energy from water droplets, as microscale energy is required for the distributed sensors in the interconnected human society. However, achieving a sustainable direct-current generating device from water flow is rarely reported, and the quantum polarization principle of the water molecular remains uncovered. Herein, we propose a dynamic water-semiconductor junction with moving water sandwiched between two semiconductors as a moving dielectric medium, which outputs a sustainable direct-current voltage of 0.3 V and current of 0.64 µA with low internal resistance of 390 kΩ. The sustainable direct-current electricity is originating from the dynamic water polarization process in water-semiconductor junction, in which water molecules are continuously polarized and depolarized driven by the mechanical force and Fermi level difference, during the movement of the water on silicon. We further demonstrated an encapsulated portable power-generating device with simple structure and continuous direct-current voltage, which exhibits its promising potential application in the field of wearable electronic generators.

**Main**

Sustainable energy harvesting from environment[1-5] is always required to meet the increasing energy demand of modern information society, especially the burgeoning Internet of Things (IoTs) and bio-electronic devices[6-9]. As a clean and abundant mechanical source, water flow energy is the most potential candidate in our daily life. For macroscale movement of water, it has been used to power the generators since the development of Faraday generator[10,11]. However, in the microscale, collecting and converting energy from flowing water molecules to electricity is still challenging, although which is indispensable for integrated generator chips with miniaturized working unit.

Actually, water is mysterious as it remains a quantum material where quantum fluctuation has a great impact on its appearance[12]. The water molecules are highly polarized, where the distribution of electron density resembles the picture of black holes[13]. Recently, water has been used to generate electricity by placing water molecular over inside carbon nanotube[14-17], graphene[18-22] or other materials[23,24]. Various physical mechanisms have been proposed to explain the mechanical energy harvesting phenomenon based on the flow of water, such as the moving liquid dragging electrons[14,21], liquid flow induced triboelectrification[24-27] and charges fluctuating asymmetric potential in substrate[15,28]. However, those mechanisms and device structure are not correlated with the abovementioned unique physical properties of water molecules. The semiconductor physics are well developed and thus the interaction between water and semiconductor could provide an insight into the electricity generation by moving water droplets on or inside various materials. Recent studies about the dynamic metal-semiconductor[29-31] and semiconductor-semiconductor interface[32-34] provide an inspiration to explore the electronic dynamics at the dynamic junction[35]. Herein, we design a dynamic water-semiconductor junction generator with water as the moving dielectric medium between two semiconductors[36], which has been rarely investigated before.

In this generator, the water droplet is sliding between a P-type silicon and a N-type silicon with different Fermi levels, which leads to a continuous current output. The physical mechanism is proposed to be originated from dynamic polarization process of moving water molecules between semiconductors. In detail, when water molecules contact with silicon, the sandwiched water molecules are instantaneously polarized and the free charge carriers in silicon are accumulated at the water-semiconductor interface, and then they reach the electrostatic balance and polarization balance, which is driven by the Fermi level difference between the water and semiconductor. As water droplet moves, the polarization balance is repeatedly broken and re-established. Meanwhile, these induced polarized electrons and holes are released and rebound to the P-type and N-type semiconductors respectively. In this way, such water-semiconductor junction with dynamic polarized water-semiconductor interface generates continuous voltage or current in the external circuit. The direct-current generator based on dynamic water-semiconductor junction with polarized water as moving dielectric medium realizes open-circuit voltage of up to 0.3 V and short-circuit current of 0.64 μA, with matched internal resistance with the traditional electronic information devices based on PN junction. We further demonstrated an encapsulated portable power generation device with simple structure and sustainable direct-current electricity. Our approach reveals the quantum polarization properties of water molecular in dynamic junction and provides a novel and promising way of converting low-frequency mechanical energy into sustainable direct-current electricity, especially the abundant water droplet energy around the world.

The three-dimensional diagram of dynamic semiconductor-liquid-semiconductor structure is shown in Figure 1a, in which the liquid droplet is located between a N-type silicon and a P-type silicon. Two polyvinyl chloride (PVC) layers with a thickness of 1 mm separate two semiconductors, which forms a small channel (Figure 1b). Liquid can move in this channel in forward or backward direction. The generated current is collected by the Ti/Au electrodes deposited on unpolished silicon surfaces in external circuit.

Those liquids whose molecular structure is asymmetric can be used in the dynamic water-semiconductor junction generator. Here, deionized water is preferred for its extreme abound and easy-availability in our daily life, such as rivers, lakes, oceans and even our bodies. In the dynamic water-semiconductor junction generator, every time a water droplet flows through the channel, a current or voltage signal is generated. As shown in Figure 1c and 1d, three peaks of voltage with maximum of 0.12 V (or current with maximum of 0.46 μA) results from three independent motion process of water droplet with volume of 50 μL at speed of 150mm/s.

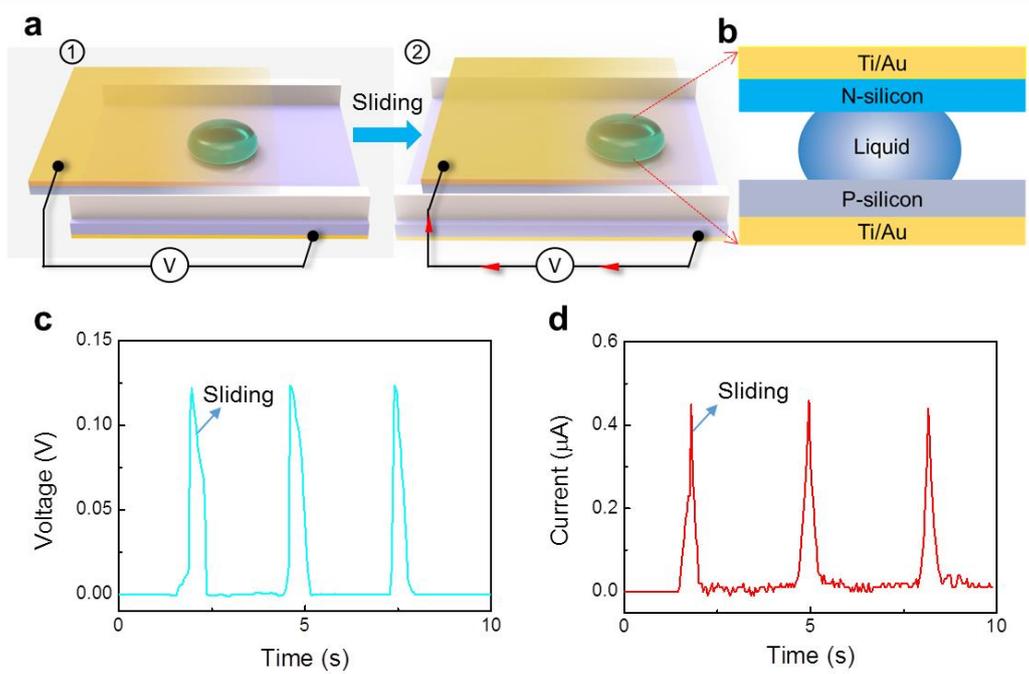

**Figure 1. The structure schematic and electricity generation of the dynamic liquid-semiconductor junction structure generator. a, the three-dimensional diagram of the dynamic liquid-semiconductor junction structure generator. b, the zoom-in schematic structure consists of two semiconductors, Ti/Au electrodes, liquid and PVC layer. c-d, the curves of the voltage generation (c), current generation (d) depends on time. Each peak of voltage and current stands for an independent process of water droplet movement.**

The origin of this electricity generation is attributed to the dynamic polarized water in the junction, which leads to the rebound or reflecting effect of the induced polarized charges. In this semiconductor-water-semiconductor structure, when water droplet contacts with two silicon wafers, the water molecules are polarized instantaneously at the water-semiconductor interface driven by the Fermi level difference between water and silicon wafer (Figure 2a-b). As water droplet moves along the semiconductors, the dynamically polarized electrons and holes at the interface are rebound to two semiconductors, respectively. In this electricity generation process, the current output particularly shows the characteristics of symmetrical output, without being restricted by the moving direction (Figure 2c). The unique isotropy of the dynamic silicon-water-silicon generator arises from the intrinsic Fermi level difference of the water and semiconductor, leading to the directional water-semiconductor polarization.

In this dynamic water-semiconductor junction, the simulation schematic diagram of polarization process is demonstrated in Figure S1 according to the physical mechanism above. The water molecules are first placed horizontally between the P-type silicon and N-type silicon. Under the Fermi level difference between the water and semiconductors, water molecules form an ordered arrangement, whose hydrogen atoms point to the N-type silicon and oxygen atoms point to the P-type silicon. After reaching the dynamic stable state, holes can be induced and accumulated in the interface of P-type silicon, and electrons can be induced and accumulated in the interface of N-type silicon, respectively. With the movement of water droplet, these interfacial carriers are released and generate directional current output. The polarized water works as a moving dielectric medium in this dynamic junction[34,36], which has a negligible damage to the semiconductor interface.

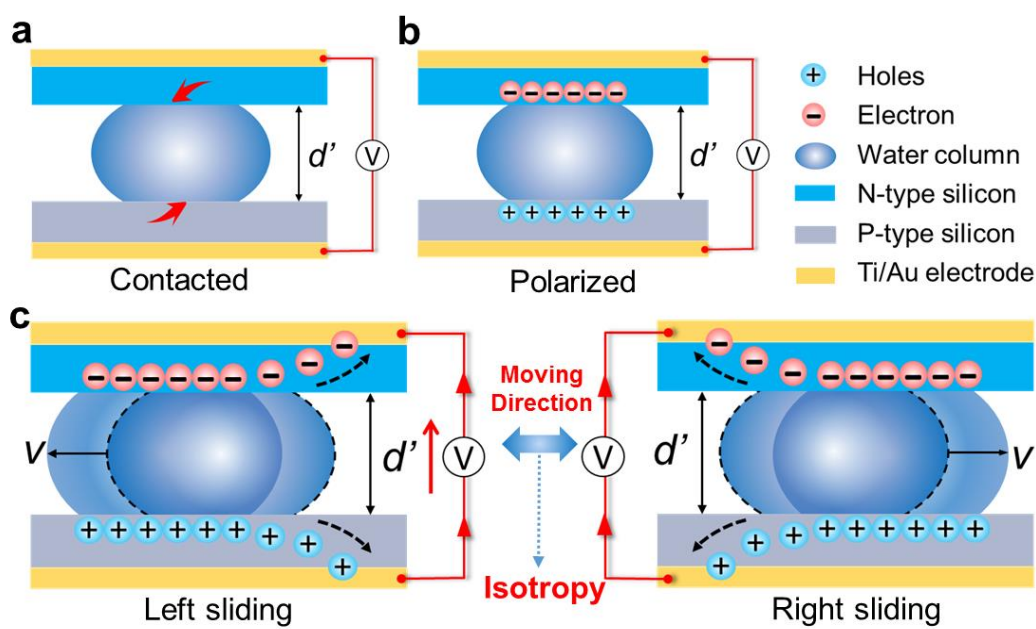

**Figure 2. The work mechanism of dynamic water-semiconductor junction generator. a-c, Schematic drawing showing the working mechanism and charge transport process of the dynamic water-semiconductor junction generator under contacted (a), polarized (b), left and right sliding (c) states. The hole and electron is represented in positive and negative sign, respectively. With the movement of water droplet, these interfacial holes and electrons are released and generate directional current output, which shows the characteristics of isotropy in the moving direction and electricity.**

According to the aforementioned mechanism, the output characteristic of generator can be determined by the dynamic polarization degree of water molecules which should be related to these parameters of speed, direction and volume. As shown in Figure 3a-d, with the increasing speed of the water (water volume of 50 μL), the output voltage (Figure 3a) and current (Figure 3b) positively increase and reach saturation values of 0.12 V and 0.46 μA respectively, at the speed up to 150 mm/s. The voltage and current curves are fitted precisely with function of $U_{oc}=0.12-0.41\times e^{-0.024v}$ and $I_{sc}=0.46-0.53/(1+e^{0.042v-1.9})$, respectively. The increase in voltage and current output is caused by the changing polarization degree of water at different moving speeds. Whether it is moving to the left or right (the silicon wafer moving speed exceeds 150 mm/s, and the water droplet volume is 50 μL), the generator always generates a positive voltage of 0.12 V from P-type silicon to N-type silicon, as shown in Figure 3c. Maintaining a constant wafer moving speed of 150 mm/s, the output current increases positively with the water droplet volume, and reaches a maximum of 0.64 μA at the volume of 150 μL (Figure 3d). The current curve can be fitted precisely with an exponential function of $I_{sc}=0.66-0.66\times e^{-0.021V}$. However, the output voltage nearly remains constant at 0.1175 ± 0.0125 V with the increase of water volume, as shown in Figure 3e. This constant voltage is caused by the limitation of the Fermi level difference between two semiconductors, which will be explored later. And the increase in output current is caused by the polarization of more water molecules per unit time. These system experiments confirm that the output of our dynamic water-semiconductor junction generator is positive related to the degree of dynamic water polarization, which is determined by its speed or volume and overcomes the limitation of the moving direction of liquid, demonstrating the proposed mechanism.

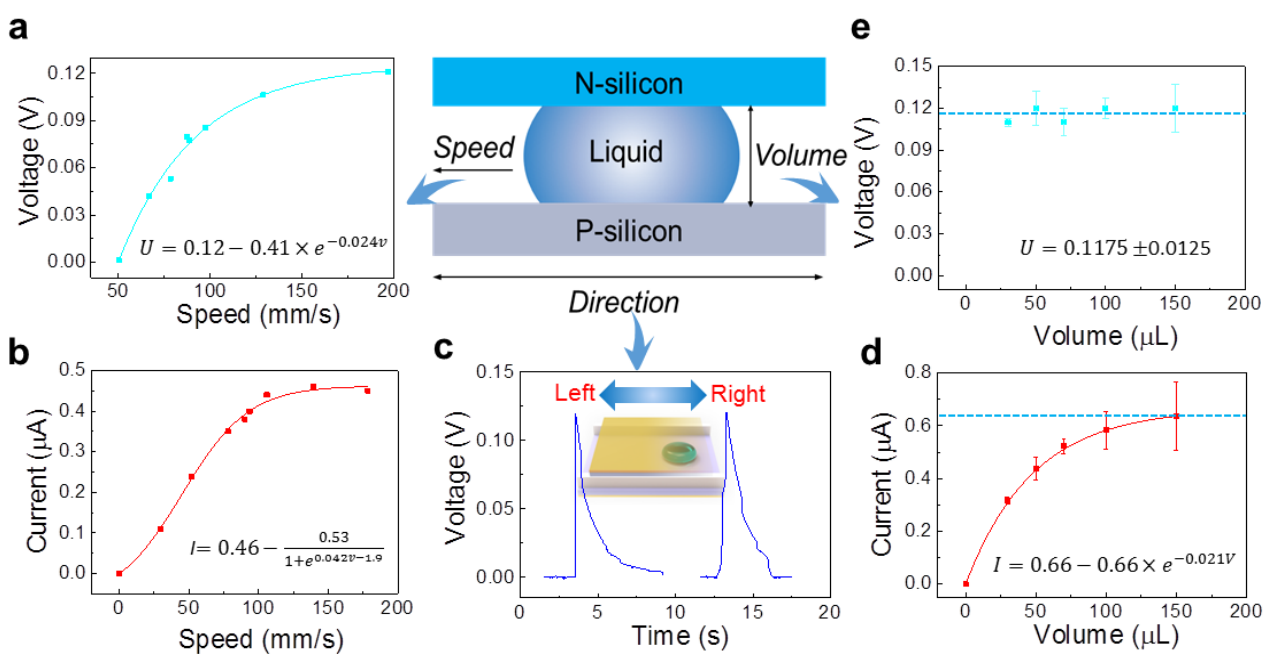

**Figure 3. The output of dynamic water-semiconductor junction generator. a-b, the output voltage (a) and current (b) depend on the speed of silicon wafer moving, on the condition of water volume of 50 μL. c, the curve of voltage output when the silicon wafer moves to the left or right at a speed exceeding 150 mm/s (water droplet volume is 50 μL). d-e, the relationship between the water volume and the output current (d) and output voltage (e), when the silicon wafer moving speed is about 150 mm/s.**

The energy band structure diagram of the static and dynamic P-Si/water/N-Si structure is shown in Figure 4a and 4b. Water molecules at water/semiconductor interface are polarized as soon as the contact of water and semiconductor (Figure 4a), under the Fermi level difference between the water and semiconductors. And the induced polarized charges make the energy band of the water/silicon surface bend. Due to the Boron or Phosphorus doping, the Fermi level of the P-type and N-type silicon is near to the valence band and conduction band, respectively. Simultaneously, the polarization also exists when the water slides along the silicon. The polarized water works as a moving dielectric medium in the dynamic junction, which can largely protect the dynamic junction from sliding wear of interface. As shown in the band diagram of dynamic structure (Figure 4b), the dynamic polarization electrons and holes in the interface are rebound to N-type and P-type semiconductors, respectively. During the dynamic polarization process, the quasi-Fermi level of the dynamic silicon-water-silicon junction is in a non-equilibrium state, which is different from the equilibrium Fermi level of the static junction in Figure 4a. As shown in Figure 4b, this potential difference and voltage output are highly correlated with the quasi-Fermi level difference of the semiconductor electrodes.

So the performance of dynamic silicon-water-silicon generators with different Fermi level differences has been measured. N-type silicon wafers with different resistivity of 0.001, 0.01, 0.5, 5, 50, 1000, 10000 Ω·cm are used, keeping the resistivity of P-type silicon unchanged at 0.001 Ω·cm in this experiment. The Fermi level of the N-type silicon can be calculated with the formula in Supporting Information. The work function of the N-type silicon with the different resistivity of 0.001, 0.01, 0.5, 5, 50, 1000 and 10000 Ω·cm is 4.13, 4.17, 4.27, 4.32, 4.38, 4.46 and 4.52 eV, respectively. In the meantime, the P-type silicon with the resistivity of 0.001 Ω·cm is used as the substrate, whose work function is 5.14 eV. So the Fermi level difference of the P-type silicon with different resistivity is as high as 1.01, 0.97, 0.87, 0.82, 0.76, 0.68 and 0.62 eV here. For the N-type silicon with the different resistivity, the output voltage of the dynamic silicon-water-silicon generator is 0.30, 0.20, 0.14, 0.12, 0.10, 0.05 and 0.01 V, respectively, under the moving speed of 150 mm/s and water volume of 50 μL. It is noteworthy that this voltage of devices with different Fermi level differences is positive related to the corresponding Fermi level differences (Figure 4c), which plays a key role in the voltage output. The detailed voltage output of dynamic silicon-water-silicon generator with different resistivity is shown in Figure S2. After inserting an insulating layer of 10 nm $HfO_2$, the output voltage is enhanced to 0.17 V due to the enhanced barrier height of the junction interface, as shown in Figure S3.

As the dynamic polarization of water molecules is the origin of power generation in dynamic silicon-water-silicon structure, different polar and nonpolar liquid are further used to test the performance of semiconductor-liquid-semiconductor structure. As shown in Figure 4d, the polar liquid including water, ethylene glycol ($(CH_3OH)_2$) and ethanol ($C_2H_5OH$) can generate voltage output of 0.12, 0.16 and 0.18 V, respectively, with the moving speed of 150 mm/s and liquid volume of 50 μL. However, the nonpolar liquid such as normal hexane ($C_6H_{14}$) cannot generate any voltage output, verifying that the dynamic polarization process of water is the key for the electricity output of dynamic semiconductor-water-semiconductor structure. In consideration of the advantages of water droplet, such as abundance and non-polluting, water is still the most potential in application.

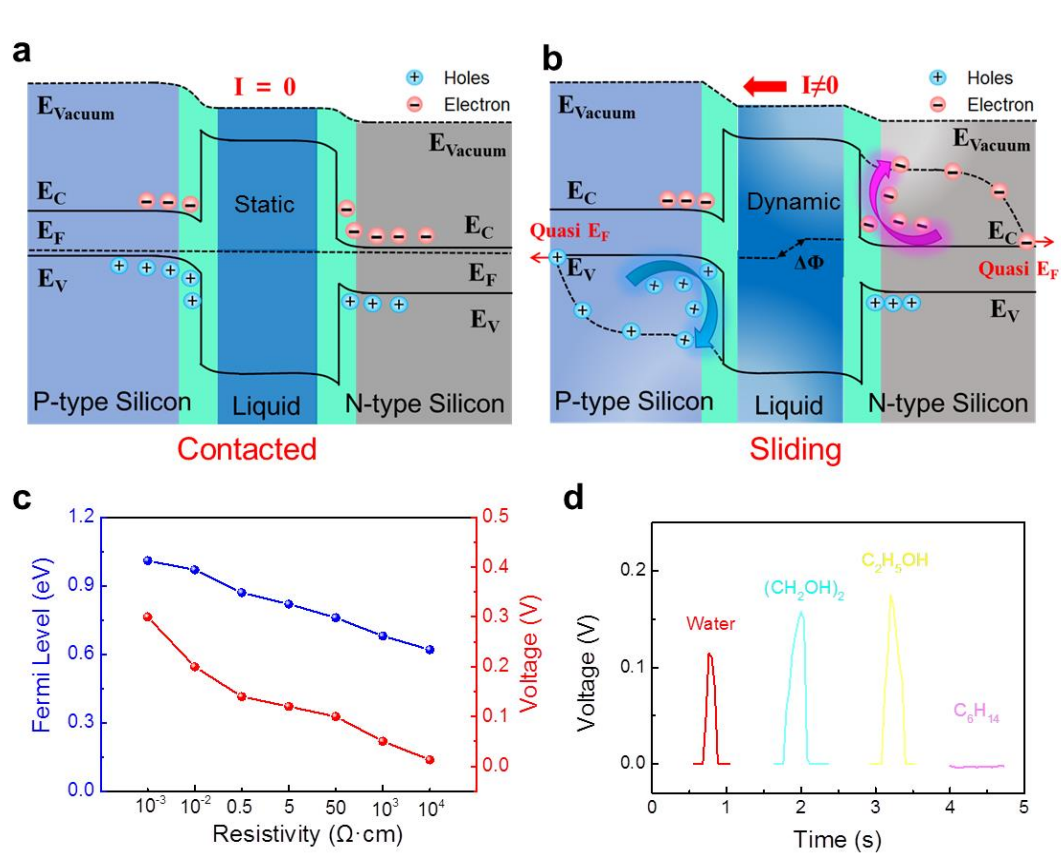

**Figure 4. The in-depth exploration about the mechanism based on Fermi level height and molecules polarization. a,** the band structure schematic and carrier kinetics of static semiconductor-liquid-semiconductor generator during contact state. **b,** the band structure schematic and carrier kinetics of dynamic semiconductor-liquid-semiconductor generator during sliding state. **c,** the relationship between the output voltage and the Fermi level difference between N-Si substrates (with different resistivity of 0.001, 0.01, 0.5, 5, 50, 1000, 10000 Ω·cm) and P-Si substrates (with constant resistivity of 0.001 Ω·cm). **d,** the output voltage of the dynamic semiconductor-liquid-semiconductor generator under different polar liquid including water, $C_2H_5OH$, $(CH_2OH)_2$ and nonpolar liquid $C_6H_{14}$.

Compared with existing water droplet generator, this dynamic liquid-semiconductor junction generator occupies the advantage of sustainable direct-current generating without the limitation of the moving direction. In order to verify the potential application of such generator, the practical work performance of the device under different load resistance has been explored. As shown in Figure 5a, the output of current and voltage of device varies with the external load resistance. With the increase of load resistance, the current output decreases from 0.44 µA to 0.08 µA and the voltage output increases from 25.77 mV to 112.12 mV, respectively. Then, the corresponding output of power can be calculated by the product of work voltage and current, as shown in Figure 5b. According to the equivalent circuit of the device, the maximum power of 22.9 µW has been achieved under the external load resistance of around 390 kΩ, indicating the internal resistance of the device is as high as 390 kΩ, which is close to the static PN junction device. This low internal-resistance property can largely reduce energy loss, indicating its potential application in the circuit unit energy supply.

Furthermore, in order to reach the practical applications requirement, a simple device demo of the semiconductor-water-semiconductor sandwich structure has been realized. A demo of this dynamic silicon-water-silicon generator is shown in Figure 5c, which consists of P-type silicon, N-type silicon wafer and water. Two silicon wafers of 2-inch diameter are glued to the inner surface of a circular plastic mold, and some water is encapsulated between two semiconductors. With the shake of the plastic mold, the water droplets slide freely between two semiconductors and generate continuous direct-current voltage as high as 0.11 V (Figure 5d). This continuity and stability voltage output with limited spikes indicates the potential prospect of the dynamic water-semiconductor junction generator in the field of wearable devices and the IoTs.

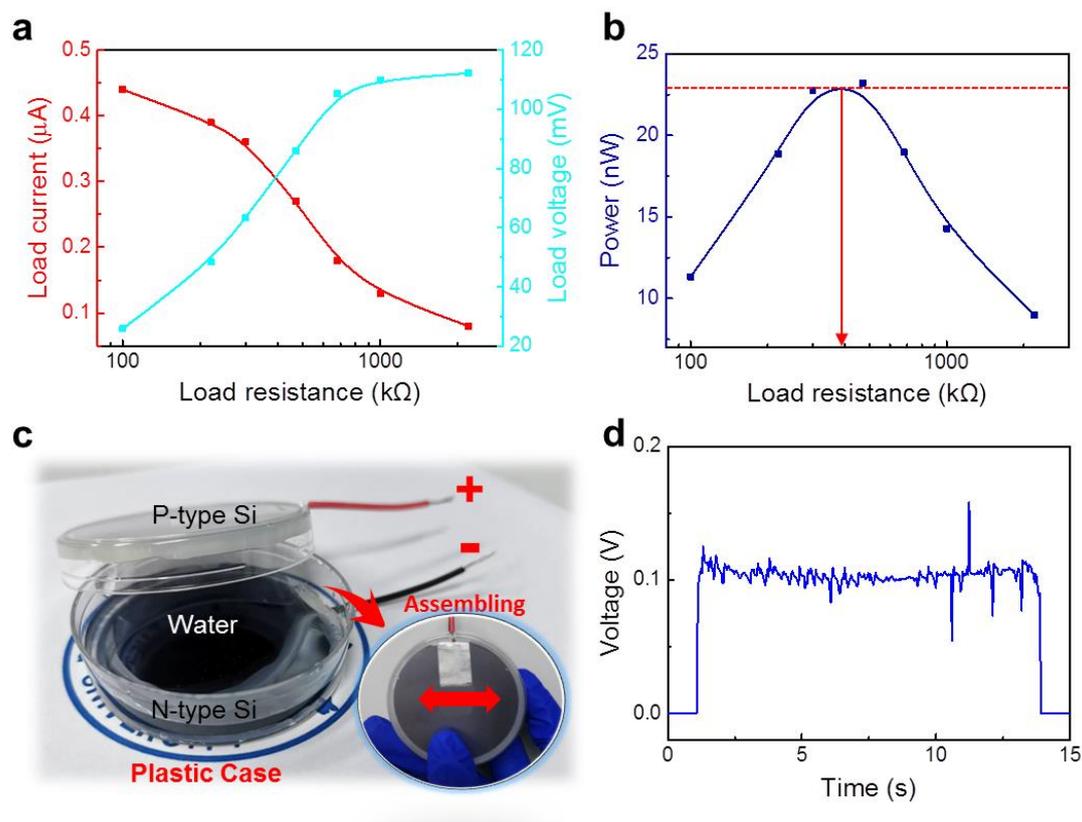

**Figure 5. The electricity output and an application demo of the dynamic water-semiconductor junction generator. a,** voltage and current output of the dynamic silicon-water-silicon generator as a function under a series of different external load resistance. **b,** power output of the dynamic silicon-water-silicon generator as a function with different load resistance. **c,** the real device picture of the dynamic silicon-water-silicon generator consists of P-type Si, N-type Silicon wafer and water. Both the semiconductors are encapsulated in plastic mold, ensuring the free movement of water between two semiconductors. **d,** the direct and continuous voltage output curve of dynamic silicon-water-silicon generator depends on time.

In this work, we have demonstrated a sustainable direct-current generator base on the dynamic semiconductor-liquid-semiconductor sandwich structure, which can harvest the energy from mechanical movement of water droplet. The mechanism can be attributed to the dynamic polarization process of water as moving dielectric medium in dynamic water-semiconductor junction. During the movement of the water droplet, the polarization balance is broken and the induced polarized charges are rebounded. Under the effect of the continuously dynamic polarized process, the induced polarized electrons and holes are rebound to N-type and P-type semiconductors respectively, forming sustainable direct-current electricity. As a representative of the generator, a voltage as high as 0.3 V and current up to 0.64 µA has been achieved based on the silicon-water-silicon structure, with matched internal resistance with the traditional PN junction-based circuit units. As a proof of principle, an encapsulated portable power-generating device with continuous direct-current voltage as high as 0.11 V has been realized, providing a novel and promising way of harvesting widely existed water droplet energy. This dynamic water-semiconductor junction generator reveals the dynamic polarization process in water-semiconductor interface and shows potential prospect in the field of wearable devices and the IoTs.

**Experimental Section**

**Preparation of Silicon Wafers:** Single polished N-type silicon wafers with different resistivity are used in the experiment, whose oxide layers are removed by being dipped into 10 wt% HF for 5 minutes. Then water, acetone and isopropyl alcohol are used to clean up the surface before the electrode fabrication. A layer of Ti with 10 nm and a layer of Au with 50 nm are grown on the unpolished side of the silicon wafer successively with a thermal evaporator, forming a natural ohmic contact electrode with silicon after annealing in 350 ℃. The same method is applied to fabricate the ohmic contact electrode of single polished P-type silicon wafer.

**Physical Measurement method:** The speed of semiconductor is explored in the electricity generating process. In this experiment, the speed is measured with an image processing analysis software called Image-Pro Plus from American MEDIA CYBERNETICS. The movement process is recorded with a high-speed camera. Keithley 2010 multimeter is used to measure the voltage and current of the generator with the sampling rate of 25 $s^{-1}$, which can be controlled by a LabView-controlled data acquisition system. We should emphasize that our output voltage and current are calculated with the average peak voltage and current in 5 points. The volume and speed can be accurately controlled with a fluctuation of 5%.

**Data availability**

The data that support the findings of this study are available from the corresponding author upon reasonable request.

# Acknowledgments

S. S. Lin thanks the support from the National Natural Science Foundation of China (No. 51202216, 51502264, 61774135 and 51991342), K. H. Liu thanks the support from Beijing Natural Science Foundation (JQ19004), Beijing Excellent Talents Training Support (2017000026833ZK11), Bureau of Industry and Information Technology of Shenzhen (No. 201901161512), Key-Area Research and Development Program of Guangdong Province (Grant No. 2019B010931001, 2020B010189001). Project funded by China Postdoctoral Science Foundation (2019M660001) and Postdoctoral Innovative Personnel Support Program (BX20180013).

## Author Contributions

S. S. Lin designed the experiments, analyzed the data, and conceived all the works. Y. H. Lu, Y. F. Yan, X. T. Yu and X. Zhou carried out the experiments, discussed the results, and wrote the paper. S. R. Feng, C. Xu, H. N. Zheng and Z. S. Yang discussed the results and assisted with experiments. L. J. Li, K. H. Liu discussed the results, analysed the data and wrote the paper. All authors contributed to the writing of the paper.

**Competing Interests:** The authors declare no competing financial interest. Readers are welcome to comment on the online version of the paper.

**Supplementary Information** is Available in the online website.